# An LLM agent for end-to-end computational materials discovery

Chen Yuntong[1], Huang Lijiang[1*], Li Yingqi[1], Liu Yu[3], Zhao Dan[4], Sun Mingqi[1], Ju Chentian[5], Liu Yanbing[3], Li Xiao-Yan[2], Zhao Guobin[2*]

1 School of Mechanical Engineering, Northwestern Polytechnical University, Xi'an, 710072, Shaanxi, China.

2* Department of Chemistry, National University of Singapore, Singapore, 117543, Singapore.

3 Institute of Information Engineering, Chinese Academy of Sciences, Beijing, 100045, China.

4 Department of Radiology, Huazhong University of Science and Technology, Wuhan, 430022, Hubei, China.

5 Department of Complexity Science and Engineering, The University of Tokyo, Kashiwa, 277-8561, Chiba, Japan

*Corresponding author(s). E-mail(s): haunglj@nwpu.edu.cn; sxmzhaogb@gmail.com;

Contributing authors: chenyt123@mail.nwpu.edu.cn; yingqili@mail.nwpu.edu.cn; liuyu@iie.ac.cn; d202382054@hust.edu.cn; sun2024@mail.nwpu.edu.cn; juct0168@gmail.com; liuyanbing@iie.ac.cn; xiaoyanli@nus.edu.sg;

## Abstract

Coordinating multi-scale tasks is an effective strategy for computational materials discovery, yet the diverse algorithms and tools involved make this process highly challenging. Here we present DREAMS, a large language model (LLM)-based agent system capable of executing the entire screening pipeline for metal-organic frameworks (MOFs). DREAMS processes a large body of MOF literature, links publications to their crystal structures, and curates them into a computation-ready database, which is then screened with progressively more expensive simulations. The promising candidates identified for separation under wet flue gas conditions all originate from independent studies. By bridging the heterogeneous stages from literature mining to high-throughput simulation, DREAMS operates across application domains and enables even non-experts to discover high-performance materials in an end-to-end, automated manner.

# 1. Main

Artificial intelligence has markedly advanced the pace of materials discovery.[1-4] A complete discovery study, however, spans stages that differ in kind, from acquiring reliable starting structures to computing their properties and screening candidates through increasingly costly methods.[5-7] How to bridge these stages into a coherent research campaign remains a central challenge.

Metal-organic frameworks (MOFs) are porous crystalline materials assembled from metal nodes and organic linkers.[8] Their tunable pore geometries and chemical environments have driven applications in gas separation,[9] catalysis,[10] and drug delivery.[11] To date, more than 100,000 experimentally determined structures have been deposited in crystallographic databases,[12] and *in silico* screening for specific applications has become a prominent research approach.[13-15] However, this requires computational chemistry researchers to master and coordinate a diverse set of skills and software tools.[16-18]

Recently, large language models (LLMs) have been applied to individual stages of this discovery process within the MOF domain,[19-23] including extracting synthesis conditions from the literature, linking material names to crystallographic identifiers, predicting adsorption properties through conversational interfaces, and repairing or generating framework structures. In parallel, curated database efforts such as the CoRE MOF 2024 have demonstrated that structure cleaning, property assignment and application screening can be integrated into a single pipeline.[24] These two lines of progress, LLM-based reasoning for individual tasks and systematic multi-stage curation, have each proved effective, but they have developed largely independently of each other.[25-27] It remains untested whether an LLM-driven agent can coordinate such a multi-stage workflow, carrying a study from a scientific question through to validated candidate materials.

Here we present DREAMS (data retrieval, extraction, and multi-scale screening), an LLM-based agent system that performs complete *in silico* screening from scratch. DREAMS processes the broad MOF literature, links publications to their crystal structures, curates the resulting collection into a computation-ready database, and screens it through methods of progressively increasing computational cost. In a case study on $CO_2$ capture from wet flue gas, DREAMS identified top-performing candidates under humid conditions, all of which originate from unrelated studies that conventional screening would be unlikely to connect. These results show that an LLM agent can carry a study from a scientific question through to validated candidates, lowering the barrier for non-experts to conduct end-to-end materials discovery.

# 2. Results

## 2.1 Overview of DREAMS

DREAMS is an LLM agent in which an orchestrator assigns tasks to four stage-specific modules (**Fig. 1 and Supplementary Video 1**), each responsible for one phase of the study: literature mining, crystal structure curation, property calculation and multi-fidelity screening. The orchestrator routes each task to the appropriate module, carries the relevant results from one phase to the next and decides at the end of each phase whether the study should proceed. Each module has its own instructions, tools and context, and all modules read from and write to a shared database. All numerical results are produced by established computational tools rather than generated by the language model. Starting from literature and structure sources, DREAMS delivers ranked candidates together with a living MOF database, which can be extended with newly deposited structures while earlier results remain unchanged.

Each module operates through an iterative plan–reflect–gate cycle. Before execution, the module examines the current data and drafts a plan that specifies tool settings, execution order and the expected range of key indicators, such as pass and error rates. After execution, it compares the observed indicators with these expectations, diagnoses any deviations and reports the results together with its diagnostics to the orchestrator, which decides whether to proceed to the next stage or halt. These judgements are made by the language model from the observed data rather than prescribed in advance, with deterministic rules as a safeguard (**Methods**), and each decision is recorded with its reasoning for auditing (**Supplementary Note 4**).

The literature module designed query terms covering variants and synonyms of MOF nomenclature and retrieved over 400,000 papers from OpenAlex.[28] It then linked these papers to crystal structures in the Cambridge Structural Database (CSD)[29] and the Crystallography Open Database (COD).[30] The two sources were highly consistent: 92.1% of publications with structures deposited in the COD also had corresponding entries in the CSD. Structures without a linked publication are not excluded from screening; instead, the literature layer provides provenance that traces each structure to its original publication, enabling the cross-domain discovery reported below. The structure module first cleaned the raw crystal structures using solvent removal tools[24,31] including symmetry reduction and solvent removal, yielding over 190,000 and 90,000 candidates from the CSD and COD, respectively. These candidates were then validated with up to five tools (Chen-Manz,[32] MOFChecker,[33] MOSAEC,[17] MOFClassifier,[16] SETC[34]). Although DREAMS supports independent execution of all tools, the tools were applied sequentially in the benchmarked order to reduce computational cost, and only structures approved by all applied tools were retained. This conservative intersection strategy yielded 64,071 and 36130 curated structures, respectively. The property module subsequently computed geometric, electronic, and chemical descriptors for these structures. Restricting both databases to CSD-derived structures with $D_f \geq 2.4$ Å deposited before 2024, DREAMS collected 99% of the computation-ready (CR) entries in CoRE MOF 2024, and most uncollected entries were not described as MOFs in their original publications. Of the collected entries, 63% were also classified as CR by DREAMS; the remainder were rejected mainly by the three additional validation tools in DREAMS (**Supplementary Note 2**).

By integrating literature mining, structure curation, featurization, and screening within a single autonomous workflow, DREAMS differs from existing systems, which each address only one or two stages of the discovery chain. To verify that the workflow supports continuous operation, we ran the complete pipeline from a single natural-language instruction on papers published during the six months following the initial database construction. DREAMS autonomously retrieved new publications, linked them to crystal structures, and processed the candidates through the same validation and featurization pipeline, incorporating 3,919 new CIFs into the database (**Supplementary Note 2.1 and Supplementary Video 2**).

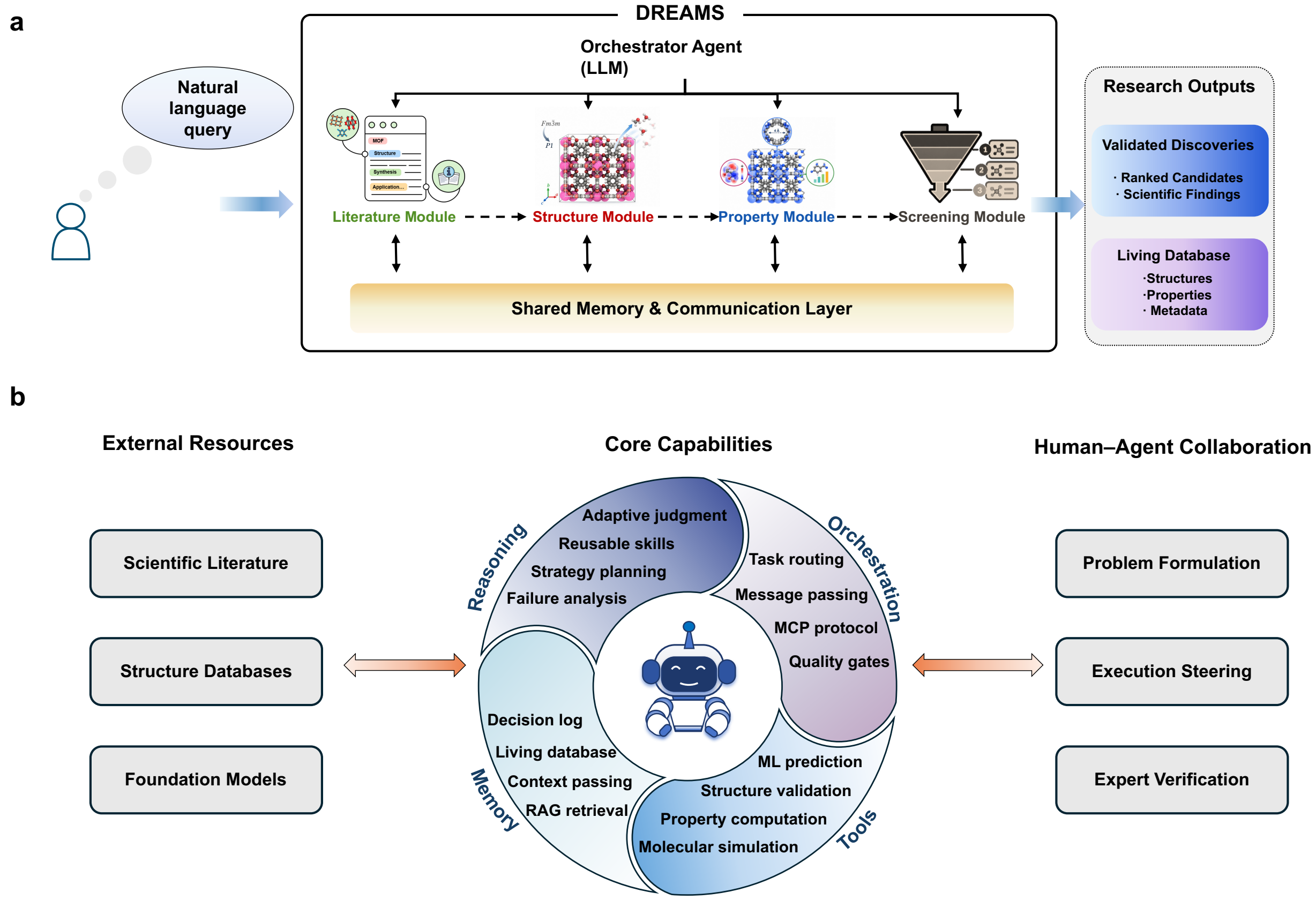

**Fig. 1 | Overview of DREAMS. a,** System architecture. Given a natural-language research question, the LLM-driven Orchestrator coordinates four stage-specific modules (Literature, Structure, Property and Screening) through a shared memory and communication layer, producing validated discoveries and a living database. **b,** Capability landscape. DREAMS draws on external resources, operates through four core capabilities, and interacts with researchers through problem formulation, execution steering and expert verification.

## 2.2 Quantitative assessment of agent capabilities

To quantitatively evaluate the performance of DREAMS, we constructed two benchmarks from the data processed in this study, covering the tracing of structures to their research origins and their validation (**Methods and Supplementary Note 4**).

To trace a structure to its research origin, DREAMS assigns the corresponding publication to one of 16 research domains. On 236 publications annotated by two researchers, all five language models were more accurate than a keyword-based classifier given the same titles and abstracts, with strict accuracies of 72–80% compared with 48% (**Fig. 2a**). The keyword classifier failed mainly on structural studies that mention a function in passing, a distinction that requires identifying the purpose of a study. Each model completed the task within minutes, compared with about 8 hours of manual annotation. These classifications underlie the domain analysis in Section 2.6.

For validation, we injected faults that raise no error message into a subset of validator result batches. DREAMS reached a balanced accuracy of 0.92, compared with 0.82 for a base agent with the same tools (**Fig. 2b**). The rule-based baseline also reran validators on many unmodified batches, whereas the LLM-only baseline missed almost all faults. The base agent, which decided for itself what to check, resolved fewer faulty batches than DREAMS, mainly those in which the faulty files were hidden among accepted ones. To check that these results were not specific to one language model, we repeated the benchmark with two others, which gave similar balanced accuracies.

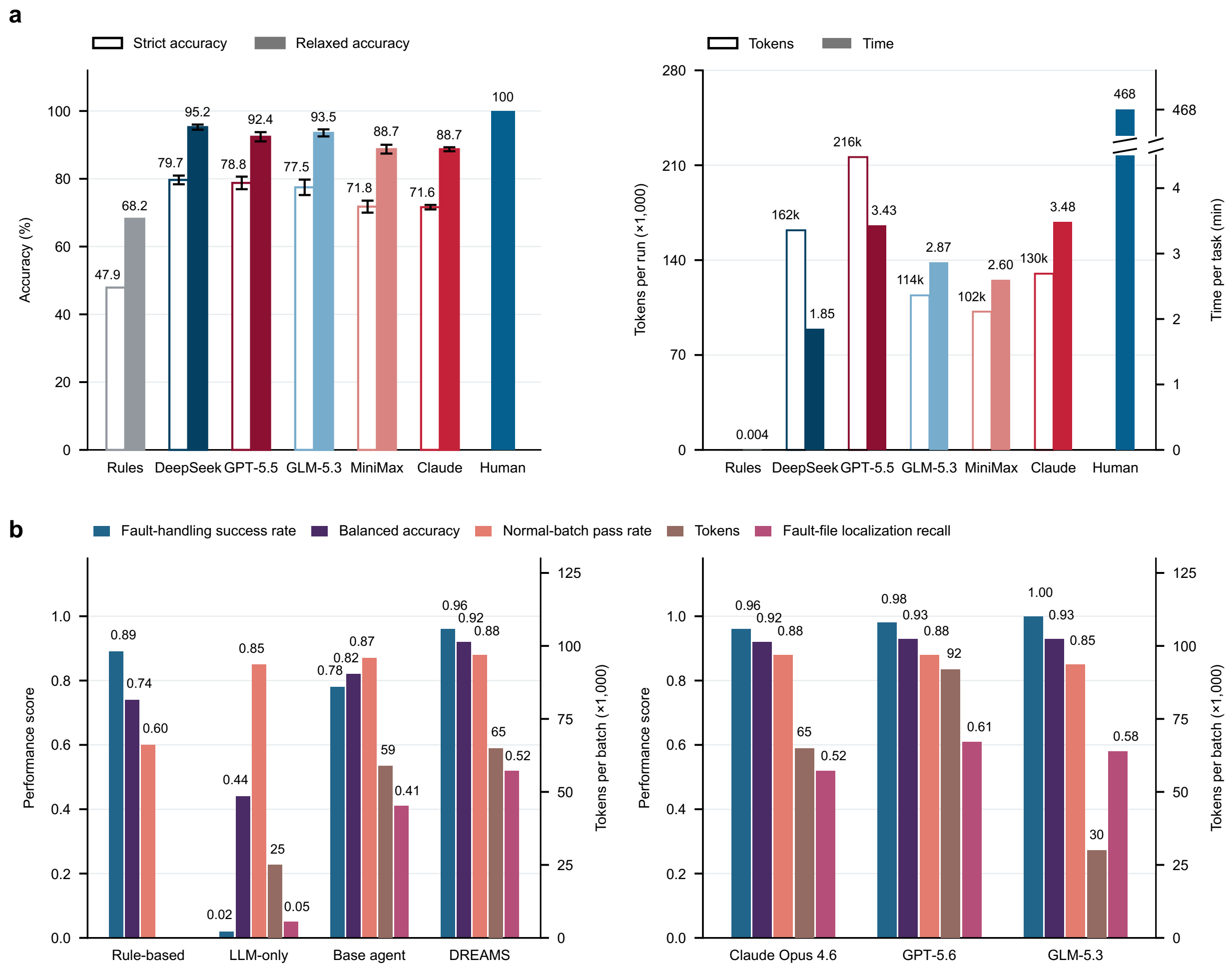

**Fig. 2 | Quantitative assessment of DREAMS on research-origin tracing and structure validation. a,** Research-origin tracing. Left, strict and relaxed accuracy of a keyword-based classifier, five language models and human annotation on 236 publications; error bars denote the standard deviation over five runs. Right, tokens per run and time per task. b, Structure validation. Left, comparison of the rule-based baseline, the LLM-only baseline, the base agent and DREAMS using Claude Opus 4.6. Right, DREAMS with three language models. Each LLM condition was run five times per batch.

## 2.3 Agent-directed screening of MOFs for $CO_2$ capture from wet flue gas

To address the objective of $CO_2$ capture from wet flue gas, the screening was designed as a four-stage workflow with specific quantitative criteria at each stage (**Fig. 3a** and **3b**). The first three stages served as filters of progressively increasing computational cost. First, rule-based prescreening removed structures with pores too small to accommodate $CO_2$, structures containing precious metals, and structures with open metal sites (OMS). Because grand canonical Monte Carlo (GCMC)[35,36] simulations insert molecules stochastically without accounting for diffusion, the results for structures with large cavities but narrow pore windows reflect a thermodynamic upper bound rather than a kinetically accessible uptake. Nevertheless, using the kinetic diameter of $CO_2$ (3.3 Å) as the cutoff would exclude promising candidates such as the rigid framework CALF-20;[37] we therefore set the pore window threshold to $D_f$ > 2.4 Å. Structures with OMS were excluded because OMS typically exhibit strong affinity for water and cannot be accurately described by generic force fields. This step reduced the number of candidates from 64,071 to 6,999.

Next, combined Widom insertion screening at 313 K imposed three criteria simultaneously: a Henry coefficient of water below $5\times10^{-6}$ mol kg$^{-1}$ Pa$^{-1}$ to ensure thermodynamic hydrophobicity,[38] a Henry coefficient of $CO_2$ above $1\times10^{-6}$ mol kg$^{-1}$ Pa$^{-1}$ to ensure sufficient affinity for the target gas, and a $CO_2/N_2$ Henry selectivity above 5. This step reduced the number of candidates from 6,999 to 638, substantially lowering the computational cost of subsequent simulations.

Finally, GCMC simulations were performed for a $CO_2/N_2$ mixture (15:85 molar ratio) at 313 K and pressures of 0.1 and 1 bar, and candidates were required to have a $CO_2$ working capacity above 0.5 mol kg$^{-1}$ and an APS above 80. CALF-20, a commercially deployed carbon capture material, was included as a benchmark, although it was excluded in the second stage owing to its Henry coefficient of water. The screening ultimately identified 11 candidates spanning two complementary regions of the selectivity–capacity design space (**Fig. 3c**).

For the 11 candidates and CALF-20, the agent computed the parameters required for vacuum pressure swing adsorption (VPSA) process simulations at 313 K. The heat capacity was calculated using the machine learning potential (MLP, uma-s-1p1, ODAC task)[39] within the harmonic approximation. Force constants were obtained by the finite displacement method implemented in Phonopy,[40,41] and thermal properties were evaluated by phonon sampling on a reciprocal-space mesh. The heats of adsorption of $CO_2$ and $N_2$ were obtained from GCMC simulations at 1 Pa using the fluctuation method implemented, approximating the infinite-dilution limit. Single-component $CO_2$ and $N_2$ adsorption isotherms were computed by GCMC simulations over 0-1 bar and fitted with the extended dual-site Langmuir model (**Eq.4** and **Eq.5**).

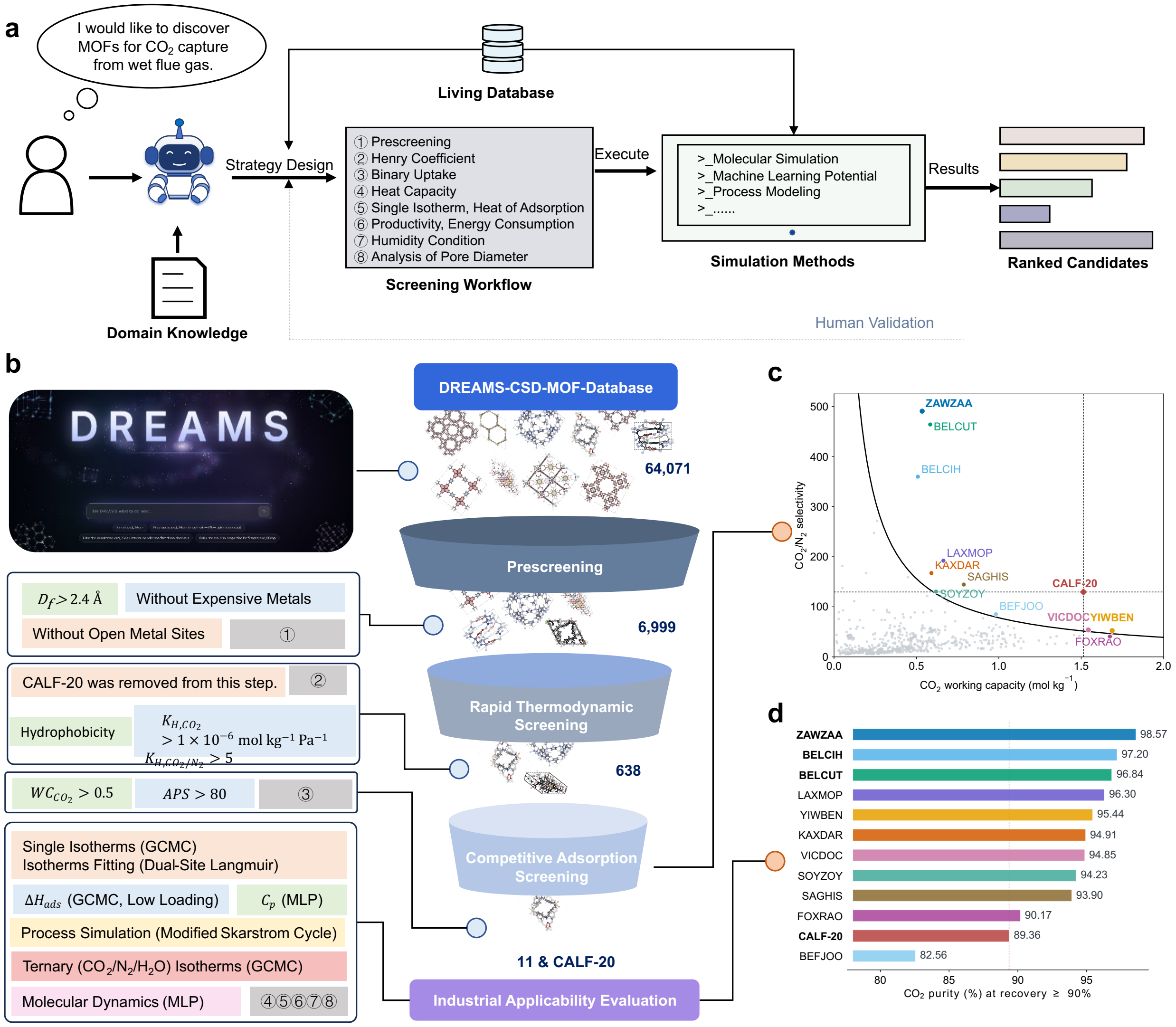

**Fig. 3 | Agent-directed multi-fidelity screening of MOFs for $CO_2$ capture from wet flue gas. a,** Architecture of the Screening Module, which translates a natural-language research objective and domain knowledge into a screening strategy, queries the living MOF database, and invokes molecular simulation, machine-learned potential and process modelling tools to return ranked candidates. The dashed arrow indicates human validation of the results. **b,** Multi-step screening funnel, showing the criteria applied at each step and the parameters calculated for the simulations. Dark blue numbers indicate the number of structures passing each step, and the circled numbers in the grey boxes correspond to the steps of the screening workflow in a. CALF-20 did not pass "Rapid Thermodynamic Screening" but was retained as a benchmark. Inset: web interface of the DREAMS database. $D_f$, diameter of the largest free sphere (equivalent to the pore limiting diameter, PLD); $K_H$, Henry coefficient; $WC$, working capacity; $APS$, adsorbent performance score; $C_p$, heat capacity; $\Delta H_{ads}$, heat of adsorption; GCMC, grand canonical Monte Carlo; MLP, machine learning potential. **c,** Results of the competitive adsorption screening. Colored points denote the 11 selected candidates and CALF-20, and grey points denote structures that were not selected. The solid black curve indicates adsorbent performance score ($APS$, **Eq. 3**) = 80, and the dashed lines indicate the $CO_2$ working capacity and $CO_2/N_2$ selectivity of CALF-20. **d,**

Process simulation results for the selected candidates, ranked by the highest $CO_2$ purity at $CO_2$ recovery ≥ 90%. The dashed red line indicates the $CO_2$ purity of CALF-20, and names in bold denote the three top-ranked candidates and CALF-20. Detailed procedures and data for each step are provided in **Supplementary Note 3**.

## 2.4 Industrial applicability of top-ranked MOFs

The VPSA simulations of a modified Skarstrom cycle evaluated two independent objectives. The process-level objective maximized $CO_2$ purity and recovery, subject to $CO_2$ recovery ≥ 90%, at fixed adsorption and desorption pressures of 1 and 0.1 bar. The economic objective maximized productivity and minimized energy consumption, subject to $CO_2$ purity and recovery ≥ 90%, with the adsorption and desorption pressures optimized within 1-10 bar and 0.1-0.5 bar, respectively. Productivity and energy consumption were used as proxies for the economic cost: lower energy consumption reduces the electricity required for capture, whereas higher productivity reduces the amount of adsorbent required. The economic ranking was determined by the compromise solution on each material's Pareto front, defined as the point with the minimum Euclidean distance to the ideal point (lowest energy consumption and highest productivity across all materials) after min-max normalization of both objectives. Under the process-level objective, ZAWZAA achieved the highest $CO_2$ purity of 98.57% (**Fig. 3d**), and 10 of the 11 candidates exceeded CALF-20. The four top-ranked materials by purity (ZAWZAA, BELCIH, BELCUT and LAXMOP) were also the four most selective materials in the competitive adsorption screening (**Fig. 3c**), indicating that $CO_2/N_2$ selectivity is a primary determinant of process-level purity. However, process performance also depends on other properties, including working capacity, framework density, heat capacity and heat of adsorption. For example, YIWBEN ranked fifth in $CO_2$ purity despite ranking 11th in selectivity, likely owing to its high $CO_2$ working capacity.

Under the economic objective, YIWBEN dominated the global Pareto front, contributing all non-dominated solutions across the 11 materials (**Fig. 4a**), and achieved both the lowest energy consumption and the highest productivity among all materials. Of the three top-ranked materials under the process-level objective (ZAWZAA, BELCIH and BELCUT), only ZAWZAA remained in the top three in the economic ranking (**Fig. 4b**). Nine of the 10 candidates outperformed CALF-20 in the economic ranking, indicating that, under dry conditions, most of the hydrophobic candidates already outperform the commercial benchmark.

Because the above evaluations were performed under dry conditions, the effect of competitive water adsorption was not considered. To verify whether the hydrophobicity criterion translates into practical moisture tolerance, we performed ternary $CO_2/N_2/H_2O$ GCMC simulations for 11 MOFs at 313 K and 1 bar across 11 relative humidity (RH) levels from 0 to 1.0. At full saturation (RH = 1.0), all 10 candidates retained over 93% of their dry $CO_2$ uptake with negligible water adsorption (**Fig. 4c**). By contrast, CALF-20 retained most of its $CO_2$ uptake up to RH ≈ 0.7, beyond which water adsorption increased sharply, and its $CO_2$ uptake nearly vanished. This result does not imply that CALF-20 is unsuitable for $CO_2$ capture, but rather that its operating window is limited to RH below approximately 0.7. This finding is consistent with previous experimental and simulation studies.[42]

Three complementary lines of evidence support the same conclusion. Widom insertion revealed low water affinity at infinite dilution, process simulations under dry conditions confirmed superior separation performance, and ternary competitive adsorption simulations demonstrated moisture tolerance at saturation. Although these methods probe different aspects of adsorption behavior, they consistently show that the candidates identified by DREAMS combine high separation performance with robust moisture tolerance. However, the rankings of individual candidates varied considerably across these metrics (**Fig. 4d**), highlighting the importance of multi-scale evaluation, which the end-to-end workflow of DREAMS enables.

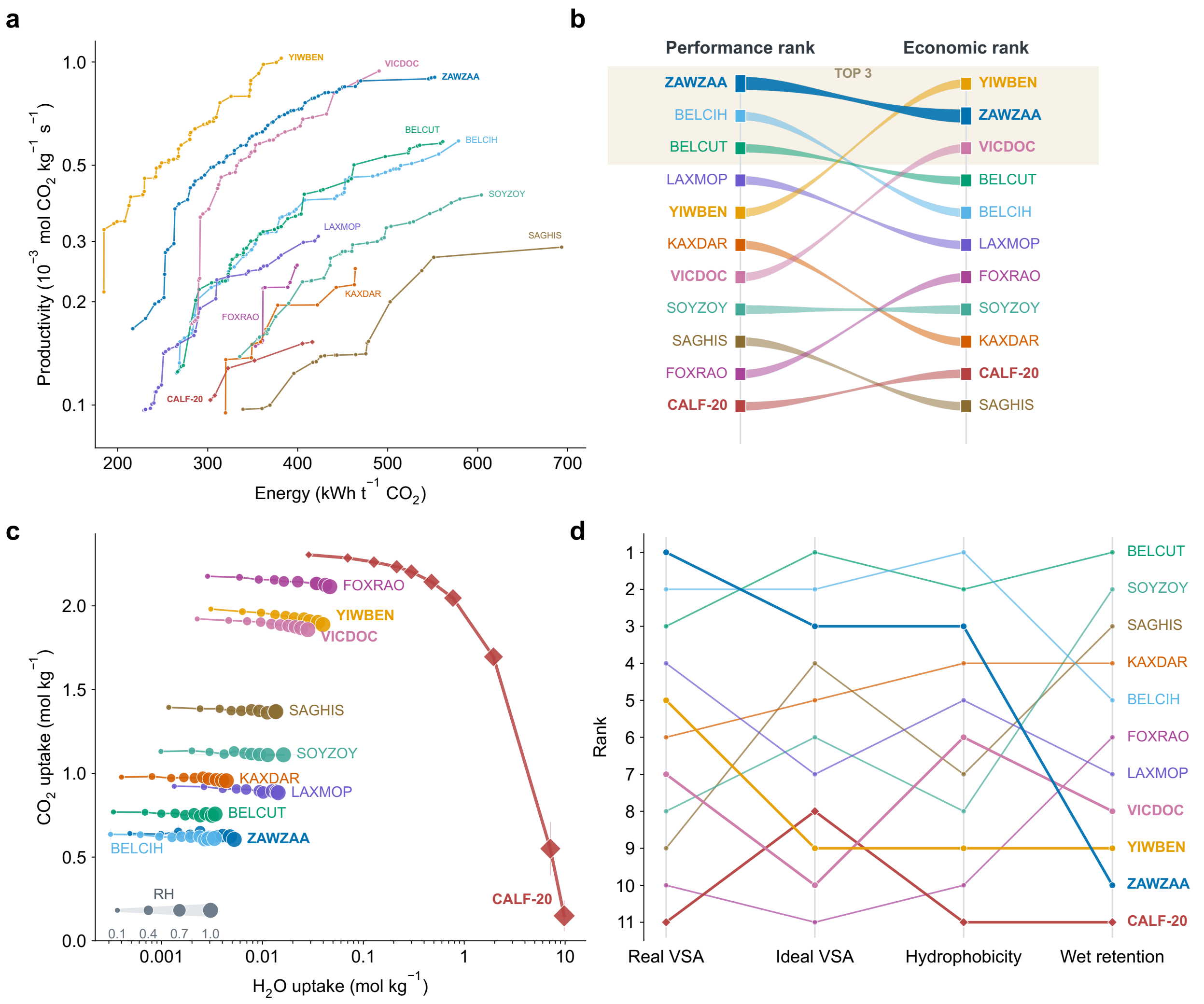

**Fig. 4 | Agent-guided industrial applicability assessment of the top-ranked MOFs. a,** Pareto fronts of $CO_2$ productivity versus energy consumption obtained from VPSA process optimization for the 10 candidates and CALF-20. **b,** Rank flow connecting the process-performance ranking, based on the highest $CO_2$ purity at $CO_2$ recovery ≥ 90%, to the economic ranking, based on the $CO_2$ capture cost at the compromise solution selected from each Pareto front. Ribbon width is for visual emphasis only and does not encode a quantitative variable. The shaded region marks the top three positions in each ranking. **c,** $CO_2$ uptake versus $H_2O$ uptake from three-component ($CO_2$/$N_2$/$H_2O$) GCMC simulations at 313 K and 1 bar with a $CO_2$/$N_2$ molar ratio of 15:85 under relative humidity (RH) from 0.1 to 1.0, with marker size indicating RH. RH = 0 serves as the dry reference and is not shown. **d,** Rankings of the 11 MOFs across four metrics: $CO_2$ purity from the real VSA (results from **Fig. 3d**) and from ideal VSA (the assumptions can be found in **Supplementary Note 3.2**), hydrophobicity based on the Henry coefficient of water at 313 K (lower values ranked higher), and wet retention, defined as the $CO_2$ uptake at RH = 1.0 relative to the dry $CO_2$ uptake. Rank 1 denotes the best performance. Names in bold denote CALF-20 and the three top-ranked candidates by $CO_2$ capture cost. BEFJOO could not meet the 90% purity and recovery targets and was excluded from the economic analysis.

## 2.5 Pore accessibility from MLP-based molecular dynamics

Because guest molecules in GCMC simulations are inserted stochastically rather than diffusing into the pores, the results cannot reveal whether a guest molecule can pass through the pore windows. DREAMS therefore used MLP-based molecular dynamics (MD) simulations to examine the fluctuations of $D_f$ and to determine whether the pore windows can accommodate the passage of a $CO_2$ molecule into the pore cavities. In each simulation, the initial configuration was first relaxed, followed by 2 ps of equilibration in the NVT ensemble and 25 ps of production in the NPT ensemble, from which snapshots were sampled every 0.1 ps (more details can be found in **Supplementary Note 3.3**). As shown in **Fig. 5a**, four conditions were simulated: the framework only, and the framework containing a single $CO_2$, $N_2$ or $H_2O$ molecule. All candidates except SOYZOY exhibited snapshots with $D_f$ exceeding 3.05 Å. For SOYZOY, $D_f$ remained below this threshold under all conditions, indicating that its pore windows may not allow the passage of a single $CO_2$ molecule, even though its pore cavities can accommodate several $CO_2$ molecules. $CO_2$ uptake in SOYZOY may therefore be kinetically limited, which is not captured by GCMC simulations. In the presence of guest molecules, the mean $D_f$ increased for more than half of the candidates, whereas only KAXDAR and SAGHIS showed a pronounced decrease. Snapshots with the smallest and largest $D_f$ (**Fig. 5b-d**) indicate that the fluctuations in $D_f$ arise mainly from linker rotation. These results demonstrate the advantage of the end-to-end screening workflow: by integrating multiple levels of simulation, DREAMS can identify potential failure modes, such as the restricted pore access in SOYZOY, that would be missed if any of these simulations were omitted.

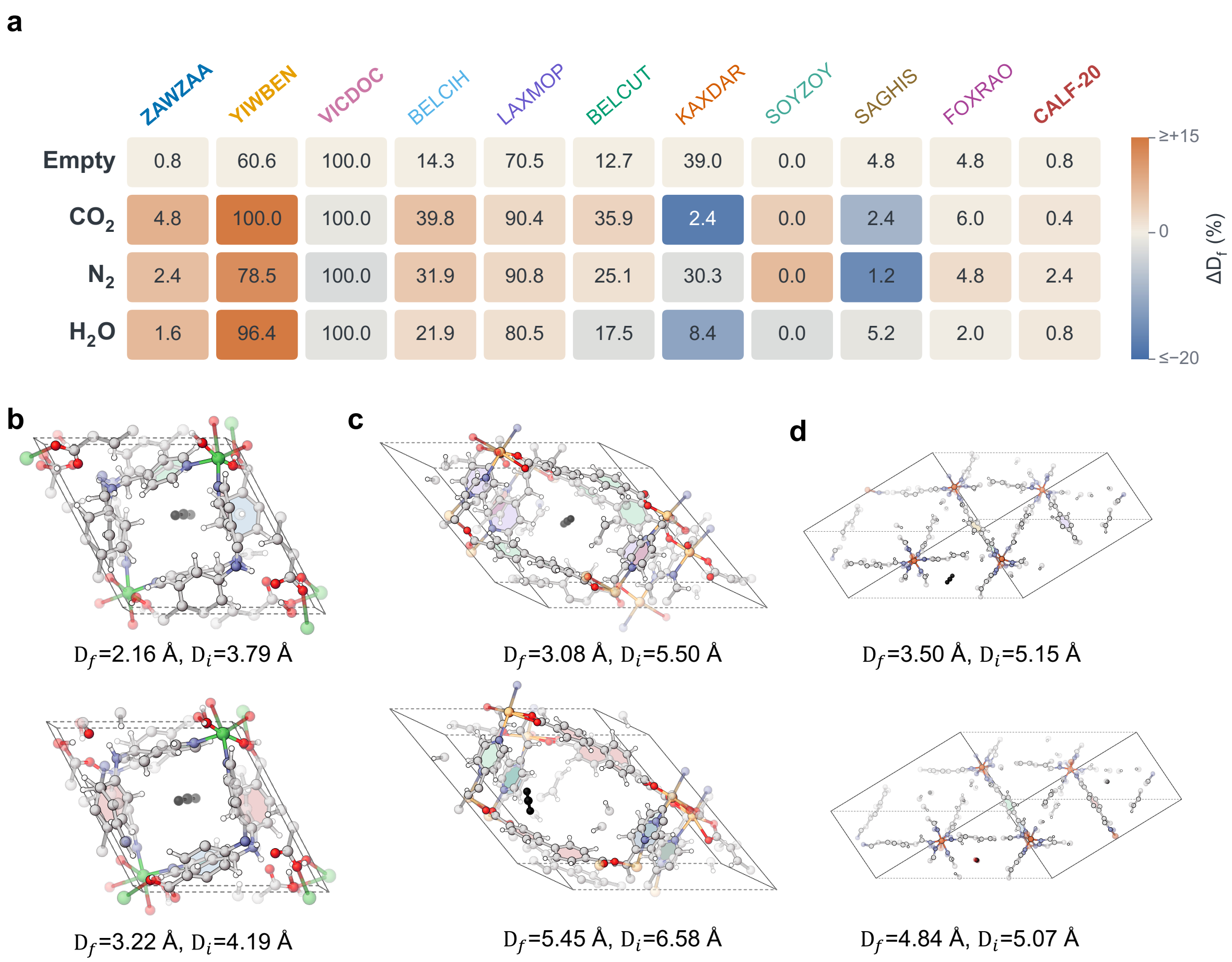

**Fig. 5 | Pore size change of the candidates from MLP-based molecular dynamics simulations performed by the screening module. a,** Mean relative change in the diameter of the $D_f$ ($\Delta D_f$) with respect to the original empty framework, for the empty framework and for frameworks containing a single CO2, N2 or H2O molecule during MD simulations at 313 K. Numbers indicate the percentage of snapshots with $D_f > 3.05$ Å, corresponding to the Lennard-Jones diameter of the oxygen atom of CO2 in the TraPPE force field.43 **b-d,** Snapshots with the smallest (top) and largest (bottom) $D_f$ for ZAWZAA, YIWBEN, and VICDOC, each containing a single CO2 molecule (shown in black). $D_i$, diameter of the largest included sphere.

## 2.6 Robust candidates emerge beyond application-driven screening

We further linked the literature information, such as the reported applications, to the corresponding structures (**Fig. 6a**). The largest source domains were luminescence and sensing, structure reports and coordination chemistry, whereas $CO_2$ capture and separation accounted for only 1.0% of structures. Nine candidates were originally reported in studies unrelated to $CO_2$ capture, and four were not included in CoRE MOF 2024 (**Fig. 6b**). ZAWZAA, the top performer under the process-level objective, was first reported in a study of luminescent Schiff-based coordination polymers, and YIWBEN, the top performer under the economic objective, originated from a study of magnetic relaxation in a Co-based MOF. None of these materials would have been identified by an application-oriented literature search for $CO_2$ capture. Using retrieval-augmented generation (RAG) over the original publications and supplementary information, DREAMS summarizes the synthesis routes and reports stability of the top-ranked candidates and integrates them with the simulated performance, with each statement linked to its source (**Fig. 6c**). This provides a direct starting point for experimental validation. These results show that an LLM-driven agent can bridge the heterogeneous stages of computational materials discovery. By exploring across application domains rather than within a single body of literature, DREAMS can uncover high-performing candidates that are easily overlooked by conventional screening workflows.

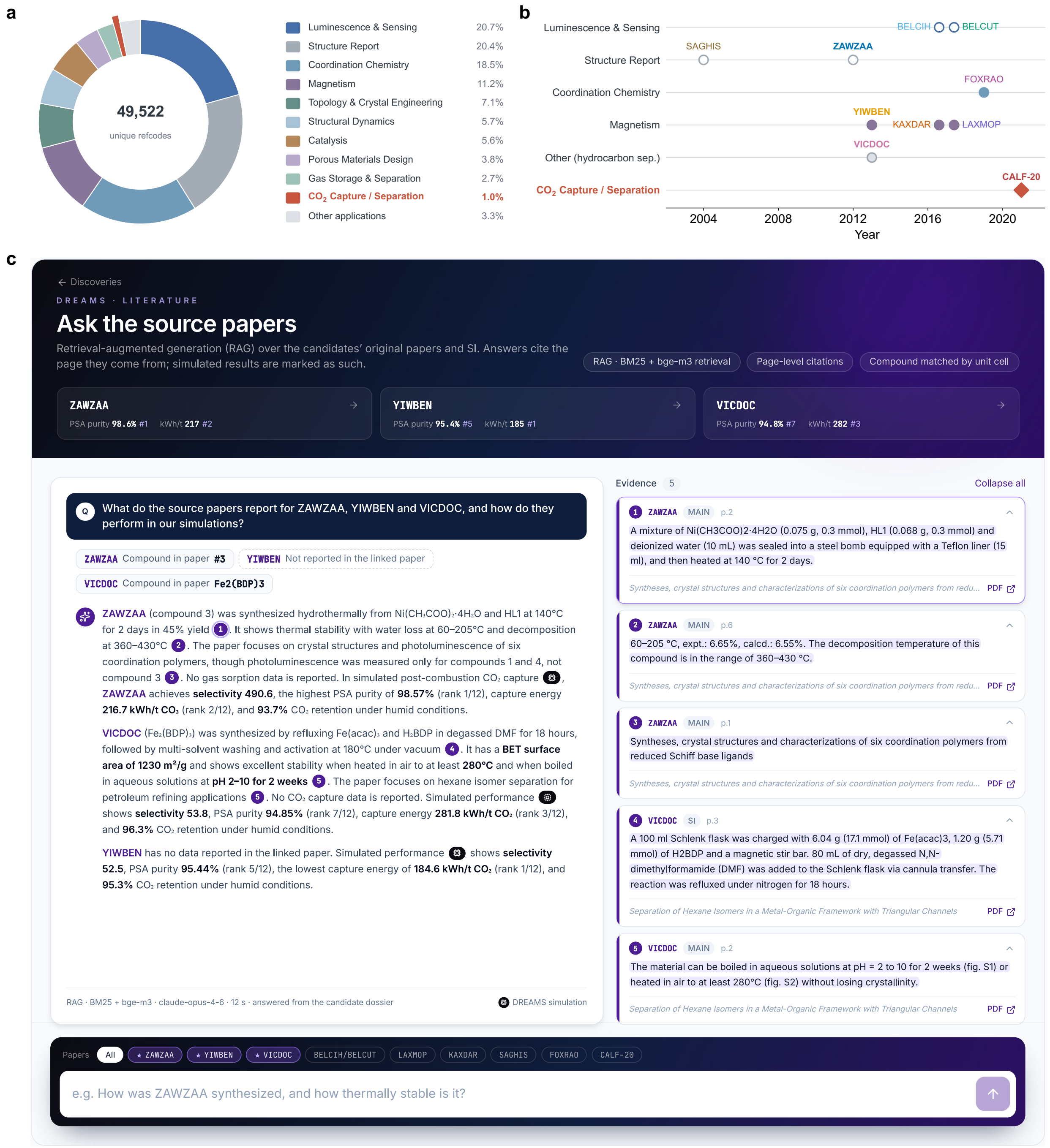

**Fig. 6 | Source domains of the curated structures and the candidates. a,** Distribution of the research domains of the original publications linked to the 49,522 unique CSD REFCODEs in DREAMS, classified by the literature module based on titles and abstracts. **b,** Research domains and publication years of the original reports of the nine candidates and CALF-20. Open circles denote candidates not included in CoRE MOF 2024, and filled circles denote candidates included in CoRE MOF 2024. Names in bold denote CALF-20 and the three top-ranked candidates in the economic ranking. **c,** Retrieval-augmented generation (RAG) interface of DREAMS, which answers queries by

combining evidence from the candidates' original publications with DREAMS simulation results. In the example shown, the source papers and simulated performance of the three top-ranked candidates are summarized.

## 3. Discussion and conclusion

This work addresses a central challenge in computational materials discovery: the stages of the discovery chain, from literature mining and structure curation to featurization and multi-scale screening, are typically performed separately and rely on manually curated databases that are updated infrequently. We developed DREAMS, an LLM agent system that performs end-to-end MOF discovery, from building a living database from scratch to identifying top-performing materials through multi-dimensional evaluation. All numerical results are produced by established computational tools, and the decisions made by the language model are recorded with their reasoning for auditing (**Supplementary Note 4**). DREAMS curated over 64,000 structures with literature provenance and incorporated newly deposited structures from a single natural-language instruction, providing a living database that can be extended with each new release. In a case study of CO2 capture from wet flue gas, the agent-directed multi-fidelity screening identified candidates that combine high process-level purity and favorable economic performance with robust moisture tolerance, retaining most of their CO2 uptake under fully saturated humidity.

Beyond automating the screening workflow, DREAMS demonstrates the value of linking structures to their literature. Most of the candidates originated from other studies rather than $CO_2$ capture, indicating that cross-domain exploration can uncover high-performing materials overlooked by application-oriented searches. The rankings of individual candidates varied substantially across adsorption, process, humidity and framework-dynamics metrics, and MLP-based MD revealed restricted pore access that GCMC simulations could not capture, underscoring the importance of multi-scale evaluation. RAG over the original publications further connects the simulated performance of each candidate with its reported synthesis and stability, providing a starting point for experimental validation.

Although the reliability of the workflow was supported by multi-tool structure validation and human validation of LLM outputs, the results should be interpreted with caution. Several aspects of this work warrant further improvement. First, the structures were not relaxed with the MLP before the phonon calculations, resulting in imaginary frequencies. Because these modes were excluded, the computed heat capacities are likely underestimated: treating the imaginary modes as real frequencies of the same magnitude or at the classical limit increases the heat capacity by 12-61% (**Supplementary Table S6**). Second, the adsorption simulations relied on force fields that may not accurately describe atomic interactions, and treated the frameworks as rigid, whereas the MD simulations indicate that linker rotation can substantially alter the $D_f$. Third, the isotherm parameters used in the process simulations were fitted at a single temperature and assumed to be temperature-independent, so that the reduction in adsorption capacity caused by bed heating was not captured. Final, the process simulations considered only a dry $CO_2/N_2$ mixture and used identical mass transfer coefficients and bed parameters for all materials. Experimental validation of the top candidates will therefore be essential before their practical deployment.

In conclusion, this study demonstrates the feasibility of an LLM agent system within a single end-to-end workflow. As a general and continuously evolving platform for materials discovery, DREAMS has the potential to accelerate the identification of high-performing MOFs and to contribute to reliable, traceable and practically relevant computational materials discovery.

# 4. Methods

## 4.1 Architecture of DREAMS

DREAMS is an LLM agent that consists of an orchestrator and four stage-specific modules, each responsible for one stage of the study. The literature module retrieves and filters publications, the structure module curates crystal structures, the property module computes structural and chemical descriptors, and the screening module performs application-specific screening. The modules do not carry out numerical calculations themselves. Instead, they call established computational tools that are registered in a common tool layer as interchangeable plug-ins, so that structure validators, solvent removal methods, charge assignment methods and property calculators can be added or replaced without modifying the agent. All modules read from and write to a shared database organized in layers. A read-only base layer preserves the data reported in this study, and newly processed data are written to a separate incremental layer, so that queries return the combined content of all layers while earlier results remain unchanged. Researchers interact with DREAMS through a web interface, a command-line interface or the Model Context Protocol (MCP), all of which access the same service layer (**Supplementary Note 4.4**).

The orchestrator converts natural-language instructions into executable tasks. For each incoming message, it first determines whether the researcher is requesting new work, revising an existing task, confirming execution, stopping a running task or asking about progress. A request for new work is decomposed into one or more tasks, each corresponding to a stage of the workflow, such as importing structures, preprocessing, structure validation, property calculation or database entry. If an instruction is ambiguous, the orchestrator asks a clarifying question before proceeding. Each task is passed to the responsible module, which drafts an execution plan as described below, and the orchestrator records the routing decision together with its reasoning. When an instruction spans several stages, the orchestrator plans the next task automatically once the previous one has finished, so that a single instruction can carry a new dataset from import to the database. After each task, the orchestrator receives the results and the diagnostic report from the module and decides whether the workflow should continue.

Each module works through a cycle of planning, execution and reflection. Before execution, the module examines the current state of the data, such as the number and composition of the input structures or publications, and drafts an execution plan that specifies the tool settings and the expected range of key indicators for each step, such as pass rates and error rates. The plan is then carried out by the computational tools without involvement of the language model. For computationally expensive stages, the tools are first run on a small number of files to detect failures before the full run begins. After each step, the agent compares the observed indicators with the expected ranges and decides whether to continue or stop, giving the reasons for its decision. Where a stage calls for closer inspection, the agent also samples its outputs for review, as in the assessment of literature filtering (**Methods 4.2.1**). At the end of each stage, the orchestrator reviews the diagnostic report and the execution statistics and decides whether to proceed to the next stage or to halt.

Each execution plan is presented to the researcher before it runs. The researcher can revise it through a form or in natural language, and all revisions are recorded with the task. Changes to thresholds require a stated reason, and a plan cannot be modified once execution has started. Deterministic rules operate alongside the judgements of the language model: for example, a property calculator is stopped if its error rate exceeds a fixed limit, regardless of the agent's assessment. Running tasks can be stopped and resumed from the last completed step.

## 4.2 Database construction and multi-scale screening

### 4.2.1 Literature retrieval and filtering

The literature module designed query terms, applied regex-based filtering rules to titles and abstracts, and verified filter quality by sampling from both the accepted and rejected pools. It estimated the filtering precision using the language model and proposed rule revisions, which were tested on a held-out sample before deciding which to adopt. The orchestrator reviewed the diagnostic report before deciding to proceed. The complete keyword list and filtering rules are provided with the code (**Code availability**).

### 4.2.2 Structure curation

Crystal structures were obtained from the CSD, linked to the collected publications through their DOIs and validated with five tools. To determine an efficient execution order, the structure module benchmarked each tool on a random sample, measuring throughput and rejection rate, and used the measurements to arrange the pipeline. All tool names, versions and configurations are provided with the code (**Code availability**).

### 4.2.3 Property calculation

The property module then computed descriptors for all computation-ready structures using nine calculators, which cover pore geometry, topology, open metal sites, metal oxidation states, revised autocorrelation descriptors, stability predictions and partial atomic charges assigned with PACMAN. Independent modules were run in parallel, and modules that depend on the outputs of others were run afterwards. Structures that failed in several modules were examined by the agent, and those with unrecoverable crystallographic files were excluded. A publicly releasable database built from the COD with the same workflow is described in **Supplementary Note 2**.

### 4.2.4 Screening workflow

Given a natural-language research objective, the screening module translated it into a multi-step workflow with quantitative criteria at each stage. The agent proposed the workflow structure, simulation conditions and candidate selection logic, which were reviewed by the researcher before execution; the prescreening criteria used in production were set by the researcher. Steps were ordered by increasing computational cost so that inexpensive filters removed the bulk of candidates before expensive simulations were applied. All simulation tools, force fields, thermodynamic conditions, screening thresholds and ranking definitions are reported in **Supplementary Note 3**.

The $CO_2$ working capacity ($WC_{CO_2}$), $CO_2/N_2$ selectivity ($S_{CO_2/N_2}$), and adsorbent performance score ($APS$) used for candidate selection were calculated as follows:

$$WC_{CO_2} = N_{CO_2,ads} - N_{CO_2 desj} \tag{1}$$

$$S_{CO_2/N_2} = \frac{N_{CO_2,ads}}{N_{N_2,ads}} \frac{y_{N_2}}{y_{CO_2}} \tag{2}$$

$$APS = WC_{CO_2} * S_{CO_2/N_2} \tag{3}$$

where $N$ is the uptake (mol $kg^{-1}$) obtained from GCMC simulations of the $CO_2/N_2$ mixture at 313 K, the superscripts $ads$ and $des$ denote the adsorption (1 bar) and desorption (0.1 bar) conditions, respectively, and $y$ is the gas-phase mole fraction ($y_{CO_2}$ = 0.15 and $y_{N_2}$ = 0.85).

### 4.2.5 Process simulation

In the process simulations, competitive adsorption was described by the extended dual-site Langmuir (DSL) model, in which $CO_2$ and $N_2$ compete for site 1, whereas site 2 adsorbs only $CO_2$

$$N_{CO_2} = \frac{q_{1,CO_2} b_{1,CO_2} P_{CO_2}}{1 + b_{1,CO_2} P_{CO_2} + b_{1,N_2} P_{N_2}} + \frac{q_{2,CO_2} b_{2,CO_2} P_{CO_2}}{1 + b_{2,CO_2} P_{CO_2}} \tag{4}$$

$$N_{N_2} = \frac{q_{1,N_2} b_{1,N_2} P_{N_2}}{1 + b_{1,CO_2} P_{CO_2} + b_{1,N_2} P_{N_2}} \tag{5}$$

where $P_{CO_2}$ and $P_{N_2}$ are the partial pressures of $CO_2$ and $N_2$. Note that he isotherm parameters were fitted and assumed to be temperature-independent in the process simulations; the heats of adsorption obtained from GCMC simulations were included only in the energy balance.

The heat capacity of each candidate was calculated within the harmonic approximation. Force constants were obtained by the finite displacement method using the unit cell, and the constant-volume heat capacity was evaluated as:

$$C_V = \frac{1}{N_q} \sum_{\mathbf{q}} \sum_{\nu=1}^{3N} k_B \left( \frac{\hbar \omega_{\mathbf{q}\nu}}{k_B T} \right)^2 \frac{\exp\left( \frac{\hbar \omega_{\mathbf{q}\nu}}{k_B T} \right)}{\left[ \exp\left( \frac{\hbar \omega_{\mathbf{q}\nu}}{k_B T} \right) - 1 \right]^2} \tag{6}$$

where the summation runs over all wave vectors $\mathbf{q}$ on a uniform sampling mesh in the first Brillouin zone and over all phonon branches $\nu$. $N_{\mathbf{q}}$ is the total number of $\mathbf{q}$-points, $N$ is the number of atoms in the primitive cell, so each $\mathbf{q}$-point has $3N$ phonon branches. $\omega_{\mathbf{q}\nu}$ is the phonon frequency of branch $\nu$ at wave vector $\mathbf{q}$. $T$ is the temperature, and $\hbar$ and $k_{\mathrm{B}}$ are the reduced Planck constant and the Boltzmann constant, respectively. Only modes with real, positive frequencies are included in the summation. The resulting heat capacity, obtained per primitive cell, is multiplied by the Avogadro constant $N_{\mathrm{A}}$ and reported in J $K^{-1}$ $mol^{-1}$.

The process-level optimization was formulated as follows and the $CO_2$ purity and recovery are defined as:

$$Purity_{CO_2} = \frac{n_{CO_2,Product}}{n_{Total,Product}} \tag{7}$$

$$Recovery_{CO_2} = \frac{n_{CO_2,Product}}{n_{CO_2,Feed}} \tag{8}$$

where $n$ denotes the number of moles of $CO_2$ or of all components in the product or feed over one cycle at cyclic steady state. The optimization problem is:

$$\begin{gathered} min. \left(1 - Purity_{CO_2}\right)^2, \left(1 - Recovery_{CO_2}\right)^2 \\ s.t. LB \leq t_{feed}, v_{feed}, r_l, r_h \leq UB \end{gathered} \tag{9}$$

where $t_{feed}$ is the feed time, $v_{feed}$ is velocity of feed, $r_l$ and $r_h$ are the light and heavy reflux ratio. The adsorption pressure, desorption pressure and column length were fixed at 1 bar, 0.1 bar and 1 m, respectively, and the column was discretized into 10 finite volumes. The optimization was performed using NSGA-II with a population size of 60 for 60 generations.

The productivity and energy consumption were defined as:

$$Productivity_{CO_2} = \frac{n_{CO_2,Product}}{m_{mat} * t_{cycle}} \quad (10)$$

$$Energy_{CO_2} = \frac{Energy_{all}}{m_{CO_2,1cycle}} \quad (11)$$

where $n_{CO_2}$,product is the amount of $CO_2$ in the product per cycle (mol), $m_{mat}$ is the mass of adsorbent in the column (kg), $t_{cycle}$ is the cycle time (s), $E_{total}$ is the total energy consumed by the compression and vacuum steps per cycle (kWh), and $m_{CO_2}$, product is the mass of $CO_2$ in the product per cycle ($1cycle$). The economic optimization problem is:

$$min. -Productivity_{CO_2}, Energy_{CO_2}$$

$$s.t. Purity_{CO_2} \geq 90\%, Recovery_{CO_2} \geq 90\%, LB \leq P_{ads}, t_{feed}, v_{feed}, r_l, r_h, P_{des} \leq UB \quad (12)$$

where $P_{ads}$ and $P_{des}$ are the adsorption and desorption pressures, varied within 1-10 bar and 0.1-0.5 bar, respectively; the bounds of the other variables were the same as in the process-level optimization. The first stage used 10 finite volumes and was initialized from the final population of the process-level optimization. Its final population then served as the starting point for a refinement stage with 30 finite volumes. Both stages used NSGA-II with a population size of 60, running for 100 and 40 generations, respectively. The $CO_2$ purity and recovery of all solutions on the final Pareto fronts were verified by re-simulation.

### 4.2.6 Retrieval of experimental reports

To connect the simulated performance of the candidates with their experimental reports, we built a retrieval-augmented generation system over the original publications, including supporting information, of the candidates and CALF-20. The documents were divided into text segments labelled with their source and page. For each query, relevant segments were retrieved by combining keyword matching (BM25) with dense embeddings (bge-m3), and were provided to the language model together with the simulation results and the correspondence between each structure and its compound label in the publication, which was verified against the reported cell parameters. The model was instructed to cite the source page for each statement, to mark simulation results separately and to state when information was not reported. Questions spanning several candidates were answered from a manually verified summary table, in which each entry is supported by a quotation from the original publication.

## 4.3 Updating the living database

To test whether the living database can be extended as new structures are deposited, the workflow was applied to structures added to the CSD during the six months following the initial database construction. A single natural-language instruction was given, and the orchestrator decomposed it into consecutive tasks covering the import of new structures and their publications, preprocessing, validation, property calculation and database entry. Each task plan was confirmed by a researcher before execution. The new data were written to the incremental layer of the database, leaving the base layer unchanged. A screen recording of the run is provided as **Supplementary Video 2.**

## 4.4 Design of agent evaluations

### 4.4.1 Literature classification

To evaluate whether an agent can assign publications to research domains from their titles and abstracts, we assembled a set of 236 MOF publications, comprising the publications linked to the high-performing structures from the screening and 100 publications sampled at random from the database. The set was shuffled and annotated by two researchers using a taxonomy of 16 research domains, five describing structural studies and eleven describing functional applications. The primary domain of each publication was annotated without knowledge of model outputs. Alternative domains were recorded during a subsequent review of disagreements between the models and the annotations (Supplementary Information). Each model was run five times, and an ensemble prediction was obtained by majority vote across models.

### 4.4.2 Fault injection

To test whether an agent can distinguish validator errors from genuinely unusual data, we constructed batches of 200 files from the production validation results of the CSD, processed in the sequence Chen–Manz, MOFChecker and MOFClassifier. The production results served as ground truth. Two types of faults were introduced. In the first, the MOFClassifier scores of 20 rejected files were replaced with scores of accepted files, so that faulty structures passed (three batches). In the second, an under-coordinated carbon flag was set for 6 or 12 accepted files, so that valid structures were rejected by MOFChecker (three batches each). Fifteen control batches contained unmodified data, including batches whose statistics resembled those of the faulty batches, such as structures that genuinely lack hydrogen atoms, and randomly drawn batches. For each count, such as the number of files a validator rejected, we took the range seen in 98% of 500 random batches as its normal range and gave these ranges to all conditions. Every faulty batch had at least one count outside its normal range. Files were split by CSD refcode into development and test pools, prompts were fixed on the development pool, and all reported results are from the test pool. Rerunning a validator was implemented by retrieving its production result, which was confirmed to match actual reruns on a subset of files.

DREAMS reflected after each validator, checking counts outside their normal ranges by rerunning a sample of the corresponding files, and a final gate reviewed all steps. It was compared with a rule-based baseline that reran any validator whose counts fell outside a range tuned on the development pool, an LLM-only baseline that received the same evidence but could not use tools, and a base agent with the same tools but without step-wise reflection or gate (Supplementary Information).

Performance was measured by balanced accuracy over faulty and unmodified batches. A faulty batch was counted as correct if all modified files were held or covered by an entire-step rerun of the faulty validator, and an unmodified batch if no file was held and no validation step was rerun (full criteria in **Supplementary Note 4.3**).

## 4.5 Language models

All language models in this study were accessed through their official APIs with a temperature of 0. Because outputs at this setting are not fully deterministic, experiments involving language models were repeated. In the production runs of DREAMS, all language-model decisions, including routing, planning, reflection and answer generation in the retrieval system, were made with Claude Opus 4.6 (Anthropic). All prompts were written in English, and researchers could give instructions in other languages. The planning prompts also include operational guidelines summarized from earlier runs (**Supplementary Information**). In the literature classification benchmark, DeepSeek-V4.1-Flash-1017, GPT-5.5, GLM-5.3, MiniMax-M3 and Claude Opus 4.6 were compared using the same prompt. In the fault-injection benchmark, all conditions were run with Claude Opus 4.6, and DREAMS was additionally run with GPT-5.6-sol and GLM-5.3 using the same prompts; each condition was run five times per batch, with responses limited to 1,500 tokens. The full prompts are provided with the code (**Code availability**).

## Data availability

The DREAMS COD dataset, including all processed crystal structures together with their metadata and computed features, is available on Zenodo at https://doi.org/10.5281/zenodo.23540448. The data from the multi-step screening and the CSD REFCODEs of the DREAMS CSD dataset are available in the GitHub repository at https://github.com/ClearPathLab/DREAMS. Owing to licensing restrictions, CSD structures are not redistributed but can be retrieved from the CSD using these REFCODEs.

## Code availability

All source code for DREAMS is available on GitHub repository at https://github.com/ClearPathLab/DREAMS. The version used in this study is archived on Zenodo at https://doi.org/10.5281/zenodo.23589225.

## Competing interests

The authors declare no competing interests.

## Contributions

Y.C. and G.Z. conceived and designed the research, developed the code and performed the calculations. L.H. and G.Z. supervised the work. Y.L., Yu L., D.Z., M.S., C.J., Yanbing. L. and X.-Y.L. assisted with the research and contributed to discussions. All authors contributed to writing and editing the paper.

## Acknowledgements

This work was supported from the National University of Singapore Start-Up Grant (NUS grant A-0010269-00-00) and the high-performance computing resources supported by NUS Information Technology (NUS-IT) in Singapore.